\documentstyle[11pt,menu99,epsfig]{article}

\begin{document}
    \setlength{\baselineskip}{2.6ex}


\font\fifteen=cmbx10 at 15pt
\font\twelve=cmbx10 at 12pt

\begin{titlepage}

\begin{center}

\renewcommand{\thefootnote}{\fnsymbol{footnote}}

{\twelve Centre de Physique Th\'eorique\footnote{
Unit\'e Propre de Recherche 7061
}, CNRS Luminy, Case 907}

{\twelve F-13288 Marseille -- Cedex 9}

\vspace{1 cm}

{\fifteen Working group summary:~{\Large{\mbox{$\pi N$}}} sigma 
term\footnote{Work supported in part by TMR, EC-contract 
No. ERBFMRX-CT980169 (EURODAPHNE)} 
\\
}

\vspace{0.3 cm}

\setcounter{footnote}{0}
\renewcommand{\thefootnote}{\arabic{footnote}}

{\bf 
Marc KNECHT\footnote{
E-mail address:~knecht@cpt.univ-mrs.fr}
}

\vspace{2,3 cm}

{\bf Abstract}

\end{center}

Several new theoretical and experimental developments concerning the 
determination of the nucleon sigma term are presented and discussed.

\vspace{6 cm}


\bigskip


\bigskip

\bigskip

\bigskip

\bigskip

\noindent
{\it Contribution to the} Eighth International Symposium on Meson-Nucleon Physics and the Structure of the Nucleon, {\it Zuoz, Engadine, Switzerland, August 15-21 1999}

\bigskip

\bigskip

\bigskip

\bigskip

\bigskip

\bigskip

\bigskip

\bigskip

\noindent December 1999

\noindent CPT-99/P.3922

\bigskip

\noindent anonymous ftp or gopher: cpt.univ-mrs.fr

\renewcommand{\thefootnote}{\fnsymbol{footnote}}

\end{titlepage}

\begin{titlepage}

$\ $

\end{titlepage}

\setcounter{footnote}{0}
\renewcommand{\thefootnote}{\arabic{footnote}}


\title{Working group summary:~$\pi N$ sigma term}
\author{M. Knecht\thanks{Work supported in part by TMR, EC-contract No. ERBFMRX-CT980169 (EURODAPHNE).} \\
{\em Centre de Physique Th\'eorique, CNRS Luminy, Case 907\\ 
F-13288 Marseille Cedex 9, France}} 

\maketitle

\begin{abstract}
\setlength{\baselineskip}{2.6ex}
Several new theoretical and experimental developments concerning the 
determination of the nucleon sigma term are presented and discussed.

\end{abstract}

\setlength{\baselineskip}{2.6ex}

\section*{INTRODUCTION}

The denomination ``sigma term'' stands, in a generic way, for the 
contribution of the quark masses $m_q$ to the mass $M_h$ of a hadronic 
state $\vert h(p)>$. 
According to the Feynman-Hellmann theorem~ \cite{FeynHell}, one has the exact 
result (the notation does not explicitly take into account the spin degrees 
of freedom)
\begin{equation}
\frac{\partial M_h^2}{\partial m_q}\,=\,<h(p)\vert ({\overline q}q)(0) 
\vert h(p)>.
\end{equation}
In practice, and in the case of the light quark flavours $q=u,d,s$, one 
tries to perform a chiral expansion of the matrix element of the scalar 
density 
appearing on the right-hand side of this formula. In the case of the pion, 
for instance, one may use soft-pion techniques to obtain the well-known 
result~ \cite{GOR} (here and in what follows, ${\cal O}(M^n)$ stands for 
corrections of order $M^n$ modulo powers of $\ln M$)
\begin{equation}\label{gmor}
\frac{\partial M_\pi^2}{\partial m_q}\,=\,-\frac{<{\overline q}q>_0}{F_0^2}
+{\cal O}(m_u,m_d,m_s)\,,
q=u,d,\ {\mbox{and}}\ \frac{\partial M_\pi^2}{\partial m_s}\,=\,
0+{\cal O}(m_u,m_d,m_s),
\end{equation}
where $<{\overline q}q>_0$ denotes the single flavour light-quark condensate 
in the $SU(3)_L\times SU(3)_R$ chiral limit, while $F_0$ stands for the 
corresponding value of the pion decay constant $F_\pi=92.4$ MeV.

In the case of the nucleon, the sigma term is defined in an analogous way, as 
the value at zero momentum transfer $\sigma\equiv\sigma (t=0)$ of the scalar 
form factor of the nucleon ($t=(p'-p)^2$, ${\hat m}\equiv (m_u+m_d)/2$),
\begin{equation}\label{defsig}
{\overline {\mbox{u}}}_N(p'){\mbox{u}}_N(p)\sigma (t) \,=\, 
\frac{1}{2M_N}<N(p')\vert {\hat m}({\overline u}u + 
{\overline d}d)(0) \vert N(p)>,
\end{equation}
and contains, in principle, information on the quark mass dependence of the 
nucleon mass $M_N$. Most theoretical evaluations of the nucleon sigma term 
consider the isospin symmetric limit $m_u=m_d$, but this is not required by 
the definition (\ref{defsig}). 

Another quantity of particular interest in this context is the 
relative amount of the nucleon mass contributed by the strange quarks of the 
sea,
\begin{equation}
y\,\equiv\,2\frac{<N(p)\vert ({\overline s}s)(0)\vert N(p)>}
{<N(p)\vert ({\overline u}u + {\overline d}d)(0) \vert N(p)>}.
\end{equation}
Large-$N_c$ considerations (Zweig rule) would lead one to expect that 
$y$ is small, not exceeding $\sim 30\%$. The ratio $y$ can be related, 
{\it via} the sigma term and the strange to non-strange quark mass ratio, 
to the nucleon matrix element of the $SU(3)_V$ breaking part of the strong 
hamiltonian, 
\begin{equation}
\sigma (1-y)\left(\frac{m_s}{\hat m}-1\right)\,=\,
\frac{1}{2M_N}<N(p')\vert (m_s-{\hat m})({\overline u}u + 
{\overline d}d - 2{\overline s}s)(0) \vert N(p)>.
\end{equation}
For the standard scenario of a strong $<{\overline q}q>_0$ condensate, 
$m_s/{\hat m}\sim 25$, 
the evaluation of the product $\sigma (1-y)$ in the chiral expansion gives 
$\sim 26$ MeV 
at order ${\cal O}(m_q)$~ \cite{GLmasses}, $\sim 35\pm 5$ MeV at order 
${\cal O}(m_q^{3/2})$~ \cite{GLmasses,gasser81}, and $\sim 36\pm 7$ MeV at 
order ${\cal O}(m_q^2)$~ \cite{borasoy}.

\section*{THE NUCLEON SIGMA TERM AND $\pi N$ SCATTERING}

Although the nucleon sigma term is a well-defined QCD 
observable, there is, unfortunately, no direct experimental access to it. 
A link with the $\pi N$ cross section (for the notation, we refer the reader 
to Refs.~ \cite{HoehlerLB,GLLS}) at the unphysical Cheng-Dashen point, 
$\Sigma \equiv F_\pi^2{\overline D}^+(\nu=0,t=2M_\pi^2)$, is furnished by a 
very old low-energy theorem~ \cite{cheng},
\begin{equation}
\Sigma = \sigma \big(1+{\cal O}(m_q^{1/2})\big).
\end{equation}
A more refined version of this statement~ \cite{brown} relates $\Sigma$ and 
the form factor $\sigma (t)$ at $t=2M_\pi^2$,
\begin{equation}
\Sigma = \sigma(2M_\pi^2) + \Delta_R ,
\end{equation}
where $\Delta_R = {\cal O}(m_q^2)$. The size of the correction $\Delta_R$, 
as estimated within the framework of Heavy Baryon Chiral Perturbation 
Theory (HBChPT), is small~ \cite{bernard1}, $\Delta_R < 2$ MeV (an earlier 
calculation to one-loop in the relativistic approach~ \cite{GSS} gave 
$\Delta_R = 0.35$ MeV).

In order to obtain information on $\sigma$ itself, one thus needs to pin 
down the difference $\Delta_\sigma\equiv\sigma(2M_\pi^2)-\sigma(0)$, and to 
perform an extrapolation of the $\pi N$ scattering data from the physical 
region $t\leq 0$ to the Cheng-Dashen point, using the existing experimental 
information and dispersion relations. 
The  analysis of Refs.~ \cite{GLS1,GLS2}, using a 
dispersive representation of the scalar form factor of the pion, gives the 
result $\Delta_\sigma = 15.2\pm 0.4$ MeV. On the other hand, from the 
subthreshold expansion
\begin{equation}\label{threxp}
{\overline D}^+(\nu=0,t) = d^+_{00} + td^+_{01} + \cdots
\end{equation}
one obtains $\Sigma = \Sigma_d + \Delta_D$, with $\Sigma_d = 
F_\pi^2(d^+_{00} + 2M_\pi^2d^+_{01})$, and $\Delta_D$ is the remainder, 
which contains the contributions from the higher order terms in the 
expansion (\ref{threxp}). In Ref.~ \cite{GLS2}, the value 
$\Delta_D = 11.9\pm 0.6$ MeV was obtained, so that the determination of 
$\sigma$ boils down to the evaluation of the subthreshold parameters 
$d^+_{00}$ and $d^+_{01}$.
Their values can in principle be obtained from experimental data on $\pi N$ 
scattering, using forward dispersion relations~ \cite{HoehlerLB,GLLS} 
\begin{equation}
d^+_{00} = {\overline D}^+(0,0) = {\overline D}^+(M_\pi,0) + {\cal J}_D(0),
\ d^+_{11} = {\overline E}^+(0,0) = {\overline E}^+(M_\pi,0) + {\cal J}_E(0),
\end{equation}
where ${\cal J}_D(0)$ and ${\cal J}_E(0)$ stand for the corresponding 
forward dispersive integrals, while the subtraction constants are expressed 
in terms of the $\pi N$ coupling constant $g_{\pi N}$ and of the S- and 
P-wave scattering lengths as follows:
\begin{equation}\label{subtract}
{\overline D}^+(M_\pi,0) = 4\pi(1+x)a^+_{0+} + 
\frac{g_{\pi N}^2x^3}{M_\pi(4-x^2)},
\ {\overline E}^+(M_\pi,0) = 6\pi(1+x)a^+_{1+} - 
\frac{g_{\pi N}^2x^2}{M_\pi(2-x)^2}.
\end{equation}
The dispersive integrals ${\cal J}_D(0)$ and ${\cal J}_E(0)$ 
are evaluated using $\pi N$ scattering data, which 
exist only above a certain energy, and their extrapolation to the low-energy 
region using dispersive methods. In the analysis of Ref.~ \cite{GLLS}, 
the two scatering lengths $a^+_{0+}$ 
and $a^+_{1+}$ are kept as free parameters of the extrapolation procedure. 
In the Karlsruhe analysis, their values were obtained from the 
iterative extrapolation procedure itself~ \cite{HoehlerLB}.
Using the partial waves of~ \cite{KochPiet,HoehlerLB}, the authors of 
Ref.~ \cite{GLLS} 
obtain the following simple representation of $d^+_{00}$ and $d^+_{01}$ (with 
$a^+_{l+}$, $l=0,1$, in units of $M_\pi^{-1-2l}$),
\begin{eqnarray}\label{dparam}
d^+_{00} &=& -1.492+14.6(a^+_{0+}+0.010)-0.4(a^+_{1+}-0.133),
\nonumber\\
d^+_{01} &=& 1.138+0.003(a^+_{0+}+0.010)+20.8(a^+_{1+}-0.133).
\end{eqnarray}
This leads then to a value $\sigma\sim 45$ MeV, corresponding to $y\sim 0.2$~
\cite{GLS1}.
Further details of this analysis can be found in Refs.~ \cite{mikko1,mikko2}.

\section*{THEORETICAL ASPECTS}

In the framework of chiral perturbation theory, the sigma term has an 
expansion of the form 
\begin{equation}\label{sigexp}
\sigma \sim \sum_{n\ge 1}\sigma_n M_\pi^{n+1}.
\end{equation}
The first two terms of this expansion were computed in the framework of the 
non-relativistic HBChPT in Ref.~ \cite{bernard2},
\begin{equation}
\sigma_1\,=\,-4c_1\,,\ \sigma_2\,=\,-\frac{9g_A^2}{64\pi F_\pi^2}.
\end{equation}
The determination of the low-energy constant $c_1$, which appears also in 
the chiral expansion of the $\pi N$ scattering amplitude, is crucial for the 
evaluation of $\sigma$. Earlier attempts, which extracted the value of $c_1$ 
from fits to the $\pi N$ amplitude extrapolated to the threshold 
region using the  phase-shifts of Refs.~ \cite{KochPiet,HoehlerLB}, 
obtained rather 
large values, $\sigma\sim 59$ MeV~ \cite{mojzis} 
($c_1=-0.94\pm 0.06$ GeV$^{-1}$), or even $\sigma\sim 70$ MeV~ \cite{fettes} 
($c_1=-1.23\pm 0.16$ GeV$^{-1}$), as compared to the result of 
Ref.~ \cite{GLS1}. 

The threshold region in the case of elastic $\pi N$ might 
however correspond to energies which are already too highy in order to make these 
determinations of $c_1$ stable as far as higher order chiral corrections are 
concerned. A new determination of $c_1$, obtained by matching the 
${\cal O}(q^3)$ HBChPT expansion of the $\pi N$ amplitude {\it inside the 
Mandelstam triangle} with the dispersive extrapolation of the data leads to a 
smaller value~ \cite{buettiker1,buettiker2}, $c_1 = -0.81\pm 0.15$ 
GeV$^{-1}$, corresponding to $\sigma\sim 40 $ MeV. It remains however to be  
checked that higher order corrections do not substancially modify this result.
Let us mention in this respect that the higher order contribution $\sigma_3$ 
(which contains a non-analytic ${\cal O}(M_\pi^4\ln M_\pi/M_N)$ piece)
in the expansion (\ref{sigexp}) has been computed in the context of the 
manifestly Lorentz-invariant baryon chiral perturbation theory in 
Ref.~ \cite{becher}, (see also~ \cite{leutwyler}). Once the expression of 
the $\pi N$ amplitude is also known with the same accuracy~ \cite{leutwyler}, 
a much better control over the chiral perturbation evaluation of $\sigma$ 
should be reached.

Finally, let us also mention that the results quoted above were based on 
the $\pi N$ phase-shifts obtained by the Karlsruhe group~ \cite{HoehlerLB}. 
Using instead the SP99 phase-shifts of the VPI/GW group, the authors of 
Ref.~ \cite{buettiker1} obtain a very different result, $c_1 \sim -3$ 
GeV$^{-1}$, which leads to $\sigma\sim 200$ MeV. Needless to say that the 
consequences of this last result ($y\sim 0.8$) would be rather difficult to 
accept. 

\section*{EXPERIMENTAL DEVELOPMENTS}

We next turn to the discussion of several new experimental results which have 
some bearing on the value of the nucleon sigma term. All numerical values 
quoted below use $M_\pi = 139.57$ MeV and $F_\pi=92.4$ MeV.

Let us start with the influence of the scattering length $a^+_{0+}$ on the 
value of the subthreshold parameter $d^+_{00}$, using Eq. (\ref{dparam}) and 
$a^+_{1+} = 0.133 M_\pi^{-3}$. 
The first line of Table 1 gives the result obtained 
from the value of the phase-shift analysis of Ref.~ \cite{HoehlerLB}. In the 
second line of Table 1, we show  the value reported 
at this conference~ \cite{leisi} and obtained  from the data on pionic 
hydrogen, $10^3M_\pi\times a^+_{0+} = 1.6\pm 1.3$.
The analysis of Loiseau {\it et al.}~ \cite{loiseau1}  
consists in extracting the combinations of scattering lengths 
$a_{\pi^-p}\pm a_{\pi^-n}$ from the value of pion deuteron scattering length 
$a_{\pi^- d}$ obtained  from the measurement of the strong interaction width 
and lifetime of the 1S level of the pionic deuterium atom~ \cite{pid1,pid2}. 
Assuming charge exchange symmetry ($a_{\pi+p} = a_{\pi^-n}$), they find 
$10^3M_\pi\times a^+_{0+} = -2\pm 1$ (third line of Table 1). 
Another determination of 
$a^+_{0+}$ is also possible using the GMO sum rule (we use here the form 
presented in~ \cite{loiseau1}, with the value of the total cross section 
dispersive integral $J^-=-1.083(25)$, expressed in mb and $a_{\pi^- p}$,  
$a^+_{0+}$ expressed in units of $M_\pi^{-1}$) 
\begin{equation}\label{gmo}
g_{\pi N}^2/4\pi = -4.50\,J^- + 103.3\,a_{\pi^- p} -103.3\,a^+_{0+}.
\end{equation}
Using the value $a_{\pi^- p} = 0.0883\pm 0.0008$ obtained by~ \cite{loiseau1} 
and the determination $g_{\pi N}= 13.51\pm 12$ from the Uppsala charge 
exchange $np$ scattering data~ \cite{loiseau2}, one obtains 
$a^+_{0+}=-0.005\pm 0.003$. 
The resulting effect on $\Sigma_d$ is 
shown on the fourth line of Table 1.

\begin{table}[htb]
\caption{$d^+_{00}$ for different values of the scattering length $a^+_{0+}$.}
\begin{center}
\begin{tabular}{c|c|c|c}
\hline\hline
       & $a^+_{0+}\times 10^3M_\pi$ & $F_\pi^2d^+_{00}$ (MeV) & $\Delta\Sigma_d$ (MeV)
\\ \hline
KH \cite{HoehlerLB}    &  $ -9.7$             & $ -91.0$   &         $0$
\\ \hline
$A_{\pi^-p}$\cite{leisi}   &     $+2\pm 1$               &  $-80\pm 1$   &      $+11$
\\ \hline
$A_{\pi^-d}$\cite{loiseau1} &  $-2\pm 1$               &  $-84\pm 1$   &      $ +7$
\\ \hline
$g_{\pi N}$\cite{loiseau2}+GMO   &     $-5\pm 3$               &  $-87\pm 3$   &      $ +4$
\\ \hline\hline
\end{tabular}
\end{center}
\end{table}

Several new determinations of the $\pi N$ coupling constant $g_{\pi N}$ have 
also been reported at this meeting, with values which differ from the 
``canonical'' value obtained long ago~ \cite{HoehlerLB}. Since most of these 
recent determinations do not result from a  complete partial-wave analysis of 
$\pi - N$ scattering data, we can only compare the effect of variations in 
the value of $g_{\pi N}$ on the subtraction terms (\ref{subtract}). 
The results are shown in Tables 2 and 3, respectively. Again, we take the value of~ \cite{HoehlerLB}  as reference point, and show the resulting changes 
for the value $g_{\pi N}= 13.73\pm 0.07$ from the 
latest VPI/GW analysis~ \cite{pavan2}. For comparison, we have also included 
the determination of~ 
\cite{loiseau1}, using the published data on the $\pi^-d$ atom~ 
\cite{pid2} combined with the GMO sum rule~ (\ref{gmo}), as well as the value 
determined from the Uppsala charge exchange $np$ scattering data~ 
\cite{loiseau2}. The repercussion on ${\overline D}^+(M_\pi,0)$ is 
negligible in all cases shown in 
Table 2, whereas in the case of ${\overline E}^+(M_\pi,0)$, the largest 
effect comes from the rather low value of $g_{\pi N}$ obtained by the VPI/GW 
analysis.

\begin{table}[htb]
\caption{The subtraction constant ${\overline D}^+(M_\pi,0)$ of 
Eq. (\protect\ref{subtract}) for different values of the $\pi N$ coupling 
constant, and for fixed value of the scattering length 
$a^+_{0+}\times 10^3M_\pi = -9.7$.}
\begin{center}
\begin{tabular}{c|c|c|c}
\hline\hline
       & $g^2_{\pi N}/4\pi$ & $F_\pi^2{\overline D}^+(M_\pi,0)$ (MeV) & $\Delta\Sigma_d$ (MeV)
\\ \hline
KH    &  $ 14.3\pm 0.2$           & $ 0.53$   &         $0$
\\ \hline
VPI/GW\cite{pavan2}   &     $13.73\pm 0.07$          &  $0.16$   &        $-0.37$
\\ \hline
$A_{\pi^-d}$+GMO\cite{loiseau1} &  $14.2\pm 0.2$     &  $0.46$   &        $-0.07$
\\ \hline
Uppsala\cite{loiseau2}   &     $14.52\pm 0.26$       &  $0.67$   &        $+0.14$
\\ \hline\hline
\end{tabular}
\end{center}
\end{table}

\begin{table}[htb]
\caption{The subtraction constant ${\overline E}^+(M_\pi,0)$ of 
Eq. (\protect\ref{subtract}) for different values of the $\pi N$ coupling 
constant, and for fixed value of the scattering length 
$a^+_{1+}\times 10^3M_\pi^3 = 133$.}
\begin{center}
\begin{tabular}{c|c|c|c}
\hline\hline
    & $g^2_{\pi N}/4\pi$ & $F_\pi^2M_\pi^2{\overline E}^+(M_\pi,0)$ (MeV) &$\Delta\Sigma_d$ (MeV)
\\ \hline
KH    &  $ 14.3\pm 0.2$           & $ 105$             &         $0$
\\ \hline
VPI/GW\cite{pavan2}   &  $13.73\pm 0.07$             &  $108$             &         $+6$
\\ \hline
$A_{\pi^-d}+GMO\cite{loiseau1}$ &  $14.2\pm 0.2$     &  $105$             &         $+1$
\\ \hline
Uppsala\cite{loiseau2}   &     $14.52\pm 0.26$     &  $104$             &         $-2$
\\ \hline\hline
\end{tabular}
\end{center}
\end{table}

Finally, we have summarized the various results in Table 4, where now the 
complete results for the determination of the dispersive integrals 
${\cal J}_D$ and ${\cal J}_E$ have beem included where possible, {\it i.e.} 
in the case of the KH~ \cite{HoehlerLB,GLLS} and of the VPI/GW~ 
\cite{pavan1,pavan2} analyses (see also Table 1 in~ \cite{pavan1}). 
The corresponding values of $\Sigma_d$ are 
given in the last column of Table 4. The analysis of the VPI/GW group 
increases the value of the sigma term by more than 25\%, as compared to the 
value extracted from the KH phase-shift analysis. This would lead to a value 
 of $y\sim 0.5$, which is rather difficult to understand 
theoretically. It should also be noticed that this large difference is due 
for a large part to the value $d^+_{01}=(1.27\pm 0.03)M_\pi^{-3}$ (including 
a shift in the value of the scattering length $a^+_{1+}$, which by itself 
accounts for half of the difference between KH and VPI/GW in the $d^+_{01}$ 
contribution in Table 4) as quoted 
by the VPI/GW group and obtained from fixed-$t$ dispersion relation. A 
similar analysis, but based on so-called interior dispersion relation 
(see for instance~ \cite{hoehler1} and references therein), yields 
a much smaller value, $d^+_{01}=1.18M_\pi^{-3}$~ \cite{stahov}, which 
{\it lowers} the VPI/GW value of $\Sigma_d$ in Table 4 by 10 MeV. It remains 
therefore difficult to assess the size 
of the error bars that should be assigned to the numbers given above. Also, 
the VPI/GW phase-shifts have sometimes been criticized as far as the 
implementation of theoretical constraints (analyticity properties) is 
concerned (see for instance~ \cite{hoehler2}). Furthermore, the issue of 
having a coherent $\pi N$ data base remains a crucial aspect of the problem.
The VPI/GW 
partial wave analyses include data posterior to the analyses of the Karlsruhe 
group, but which are not always mutually consistent (see {\it e.g.}~ 
\cite{mikko2} and references therein). 
Hopefully, new experiments (see~ \cite{smith}), 
will help in solving the existing discrepancies.

\begin{table}[ht]
\caption{Comparison of the values of the subthreshold parameters $d^+_{00}$ and 
$d^+_{01}$ according to differences in the input discussed in the text.}
\begin{center}
\begin{tabular}{c|c|c|c|c}
\hline\hline
      &  $F_\pi^2d^+_{00}$ (MeV)  &$2M_\pi^2F_\pi^2d^+_{01}$ (MeV)  &  $\Sigma_d$ (MeV)  
\\ \hline
KH    &  -89.4             &     139.2                &      50  
\\ \hline
VIP/GW\cite{pavan1}  &  -77.3             &     155.2                &      50 +12+16 
\\ \hline
$A_{\pi^-d}+GMO$\cite{loiseau1} & -83              &      $-$                   &      50+6 
\\ \hline
Uppsala\cite{loiseau2} & -86              &      $-$                   &      50+3.5   
\\ \hline\hline
\end{tabular}
\end{center}
\end{table}

Finally, it should be  stressed that the above discussion is by no means  
a substitute for a more elaborate analysis, along the lines of 
Ref.~ \cite{GLLS}, 
for instance (see also~ \cite{hoehler1} and~ \cite{pavan1}). 
Such a  task would have been far beyond the competences of the present 
author, at least within a reasonable amount of time and of work. 
Nevertheless,  very useful discussions with  G. H\"ohler, M. Pavan, 
M. Sainio and J. Stahov greatly improved the author's understanding of 
this delicate subject. The author also thanks R. Badertscher and the 
organizing committee for this very pleasant and lively meeting in Zuoz.

\bibliographystyle{unsrt}

\end{document}